\documentclass[aps,showpacs,nofootinbib,superscriptaddress]{revtex4}
\usepackage{epsf}
\usepackage{graphicx}
\usepackage{amsmath}
\def\slashchar#1{\setbox0=\hbox{$#1$}
   \dimen0=\wd0 \setbox1=\hbox{/} \dimen1=\wd1
   \ifdim\dimen0>\dimen1 \rlap{\hbox to \dimen0{\hfil/\hfil}} #1
   \else  \rlap{\hbox to \dimen1{\hfil$#1$\hfil}} / \fi}

\newcommand{\be}{\begin{equation}}
\newcommand{\ee}{\end{equation}}
\newcommand{\bea}{\begin{eqnarray}}
\newcommand{\eea}{\end{eqnarray}}

\begin{document}

\title{Heavy Quark Spin Symmetry and Heavy Baryons: Electroweak
  Decays\footnote{Presented at the 21st European Conference on
  Few-Body Problems in Physics, Salamanca, Spain, 30 August - 3
  September 2010.}}

\author{ C.Albertus} \affiliation{Departamento de F\'\i sica
Fundamental, Universidad de Salamanca, E-37008 Salamanca, Spain}
\author{J.E. Amaro}\affiliation{Departamento de F{\'\i}sica At\'omica,
Molecular y Nuclear, Universidad de Granada, E-18071 Granada, Spain}
\author{E. Hern\'andez} \affiliation{Departamento de F\'\i sica
Fundamental e IUFFyM,\\ Universidad de Salamanca, E-37008 Salamanca,
Spain} \author{J.~Nieves} \affiliation{Instituto de F\'\i sica
Corpuscular (IFIC), Centro Mixto CSIC-Universidad de Valencia,
Institutos de Investigaci\'on de Paterna, Aptd. 22085, E-46071
Valencia, Spain}


\begin{abstract}
Heavy quark spin symmetry is discussed in the context of single and
doubly heavy baryons. A special attention is paid to
the constraints/simplifications that this symmetry imposes on the
non-relativistic constituent quark model wave functions and on the
$b\to c$ semileptonic  decays of these hadrons.


\end{abstract}

\maketitle
\section{Introduction}
\label{intro}
Heavy Quark Spin Symmetry (HQSS) has proved to be a useful tool to
understand the bottom and charm physics~\cite{IW89,Ne94}, and it has
been extensively used to describe the dynamics of systems containing a
heavy quark $c$ or $b$. HQSS is an approximate symmetry of Quantum
Chromodynamics (QCD), which appears in systems containing heavy quarks
with masses ($m_Q$) that are much larger than the typical quantities
($q_{\rm light}=\Lambda_{QCD}$, $m_u,m_d,m_s\cdots$ ) that set up the
energy scale of the dynamics of the remaining (light) degrees of
freedom.  HQSS predicts that all type of spin interactions vanish for
infinitely massive quarks: the dynamics is unchanged under arbitrary
transformations on the spin of the heavy quark ($Q$). The
spin-dependent interactions are proportional to the chromomagnetic moment
of the heavy quark, and so are of the order of $1/m_Q$. The total
angular momentum of the hadron $\vec{J}$ is a conserved operator, the
spin of the heavy quark $\vec{S}_Q$ is conserved in the $m_Q\to
\infty$ limit, and therefore the spin of the light degrees of freedom 
$\vec{S}_l=\vec{J}-\vec{S}_Q$ is also conserved in the heavy quark
limit. Heavy hadrons come in doublets (unless $s_l=0$, with
$\vec{S}_l^2=s_l(s_l+1)$) containing states with total spin
$J_\pm=s_l\pm 1/2$ obtained by combining the spin of the light degrees
of freedom with the spin of the heavy quark $s_Q=1/2$. These doublets
are degenerated in the $m_Q\to \infty$ limit. HQSS has not been
systematically employed in the context of non-relativistic constituent
quark models (NRCQM's). NRCQM's, based upon simple quark-quark
potentials, partially inspired by QCD, lead to reasonably good
descriptions of hadrons as bound states of constituent quarks. Most of
the quark-quark interactions include a term with a shape and a color
structure determined from the one gluon exchange contribution and a
confinement potential, which is assumed to come from the long-range
nonperturbative features of QCD.

In this contribution, we will study masses and electroweak decays of
various single and double heavy baryon within a NRCQM scheme, but taking
explicitly into account HQSS constraints, which, as we will show, lead to 
great simplifications in these systems.

\section{Single Heavy Baryons}
\label{sec:1}
 Up to corrections of the order ${\cal O}(q_{\rm light}/m_Q)$, HQSS
 guaranties that the heavy baryon light degrees of freedom quantum
 numbers, compiled in Table~\ref{tab:summ}, are always well
 defined. The symmetry also predicts that the baryon pair $\Sigma,
 \Sigma^*$ (or the $\Xi^\prime, \Xi^*$ pair, or the $\Omega,\Omega^*$
 one) become degenerated for an infinitely massive heavy quark, since
 both baryons have the same cloud of light degrees of freedom.  In
 this section, we
 first describe a variational approach~\cite{Albertus:2003sx} for the
 solution of the non-relativistic three-body problem in baryons with a
 heavy quark. Thanks to HQSS, the proposed method turns out to be
 quite simple, leads to simple and manageable wave functions and
 reproduces previous results (baryon masses, charge and mass radii,
 $\cdots$) obtained by solving the Faddeev
 equations~\cite{SilvestreBrac:1996bg}. We will also discuss in this
 context, the semileptonic decay $\Lambda_b \to \Lambda_c^+ l^-
 \bar\nu_l$ \cite{Albertus:2004wj}.
\begin{table}
\caption{Quantum numbers, experimental and lattice QCD
masses of the baryons containing a single heavy quark. $I$, and
$S_{\rm light}^\pi$ are the isospin, and the spin parity of the light
degrees of freedom and $S$, $J^P$ are strangeness and the spin parity
of the baryon ($l=u,d$). } 
\label{tab:summ}  
\begin{center}
\begin{tabular}{lclllcll}\hline\noalign{\smallskip}
Baryon &$S$&$J^P$&$I$&$S_{\rm light}^\pi$& Quark content 
& $M_{exp.}$ \cite{Amsler:2008zz}& $M_{Latt.}$~\cite{bowler}
\\
       &       &         &   &          &               & [MeV]  & [MeV]   
\\\noalign{\smallskip}\hline\noalign{\smallskip}
$\Lambda_c$& 0 &$\frac12^+$& 0 &$0^+$&$udc$& $2286.46 \pm 0.14$ & $2270 \pm 50  $
\\
$\Sigma_c$ & 0 &$\frac12^+$& 1 &$1^+$&$llc$&$2453.5 \pm 0.5$  & $2460 \pm 80 $
\\
$\Sigma^*_c$ & 0 &$\frac32^+$& 1 &$1^+$&$llc$& $2517.5 \pm 0.5$ &$ 2440
\pm 70 $
\\
$\Xi_c$ & $-$1 &$\frac12^+$&$\frac12$&$0^+$&$lsc$& $2469.5 \pm 1.5$ & $2410
\pm 50 $
\\
$\Xi'_c$ & $-$1 &$\frac12^+$&$\frac12$&$1^+$&$lsc$& $2577 \pm 3$ & $2570
\pm 80$ 
\\
$\Xi^*_c$ &$-$1&$\frac32^+$&$\frac12$&$1^+$&$lsc$& $2645.9 \pm 0.6$ & $2550
\pm 80$
\\
$\Omega_c$ &$-$2 &$\frac12^+$& 0 &$1^+$&$ssc$& $2695.2 \pm 1.7$ & $2680 \pm 70$
\\
$\Omega^*_c$ &$-$2 &$\frac32^+$& 0 &$1^+$&$ssc$&   $2765.9 \pm 2.0$  & $2660 \pm 80$
\\\noalign{\smallskip}\hline\noalign{\smallskip}
$\Lambda_b$& 0 &$\frac12^+$& 0 &$0^+$&$udb$& $ 5620.2 \pm 1.6$  & $5640
\pm 60 $
\\
$\Sigma_b$ & 0 &$\frac12^+$& 1 &$1^+$&$llb$& $5811 \pm 4$  & $5770 \pm 70 $
\\
$\Sigma^*_b$ & 0 &$\frac32^+$& 1 &$1^+$&$llb$& $ 5832 \pm 4$ & $5780 \pm 70$
\\
$\Xi_b$ & $-$1 &$\frac12^+$&$\frac12$&$0^+$&$lsb$& $ 5790.5 \pm 2.7$   & $5760 \pm 60 $
\\
$\Xi'_b$ & $-$1 &$\frac12^+$&$\frac12$&$1^+$&$lsb$&    & $5900 \pm 70$
\\
$\Xi^*_b$ &$-$1&$\frac32^+$&$\frac12$&$1^+$&$lsb$&  & $5900 \pm 80 $
\\
$\Omega_b$ &$-$2 &$\frac12^+$& 0 &$1^+$&$ssb$& $ 6071 \pm 40$  & $5990 \pm 70 $
\\
$\Omega^*_b$ &$-$2 &$\frac32^+$& 0 &$1^+$&$ssb$&     & $6000 \pm 70$\\
\noalign{\smallskip}\hline
\end{tabular}
\end{center}
\end{table}

\begin{figure}
\begin{center}
\makebox[0pt]{\includegraphics[width=0.33\textwidth]{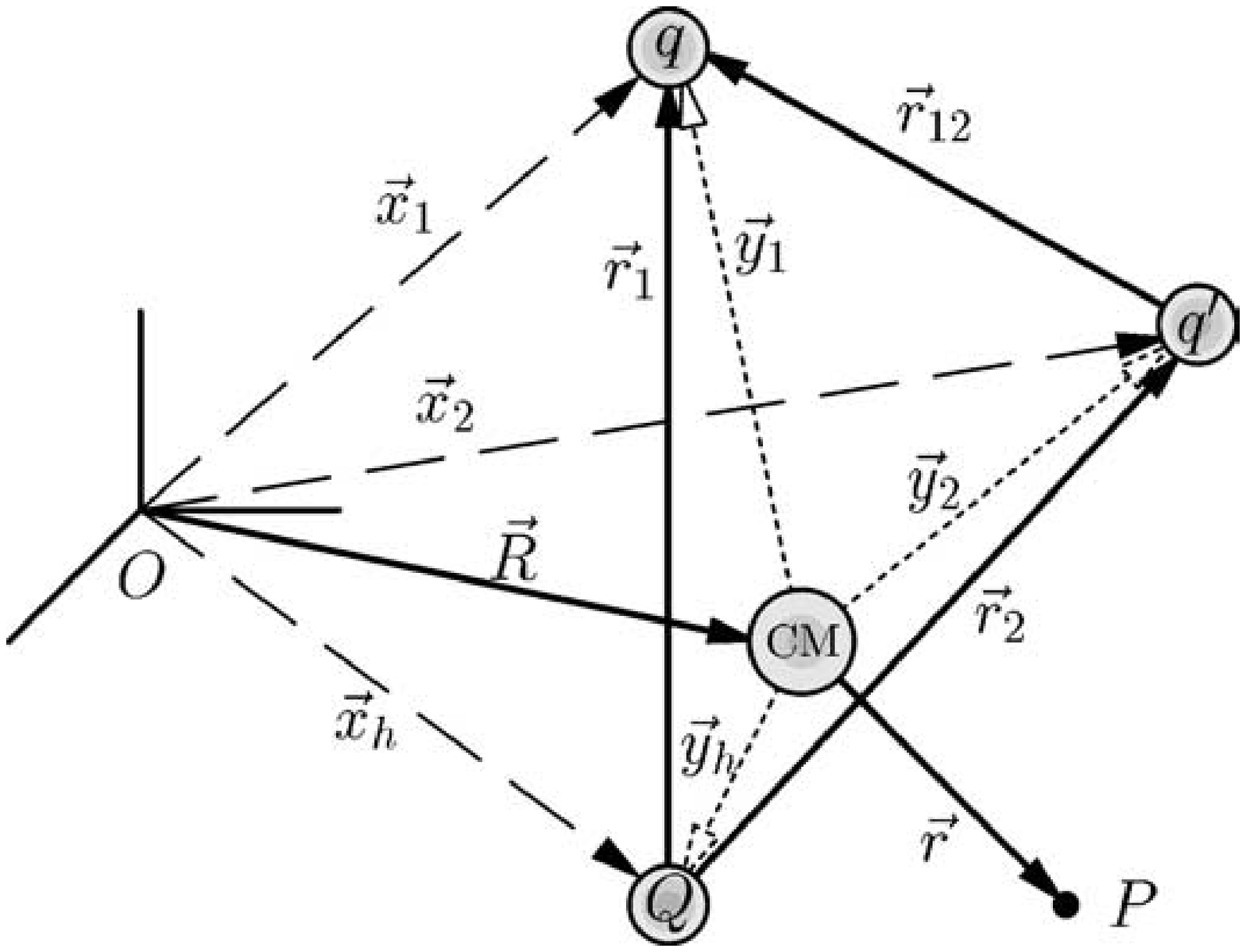}\hspace{1cm}\includegraphics[width=0.15\textwidth]{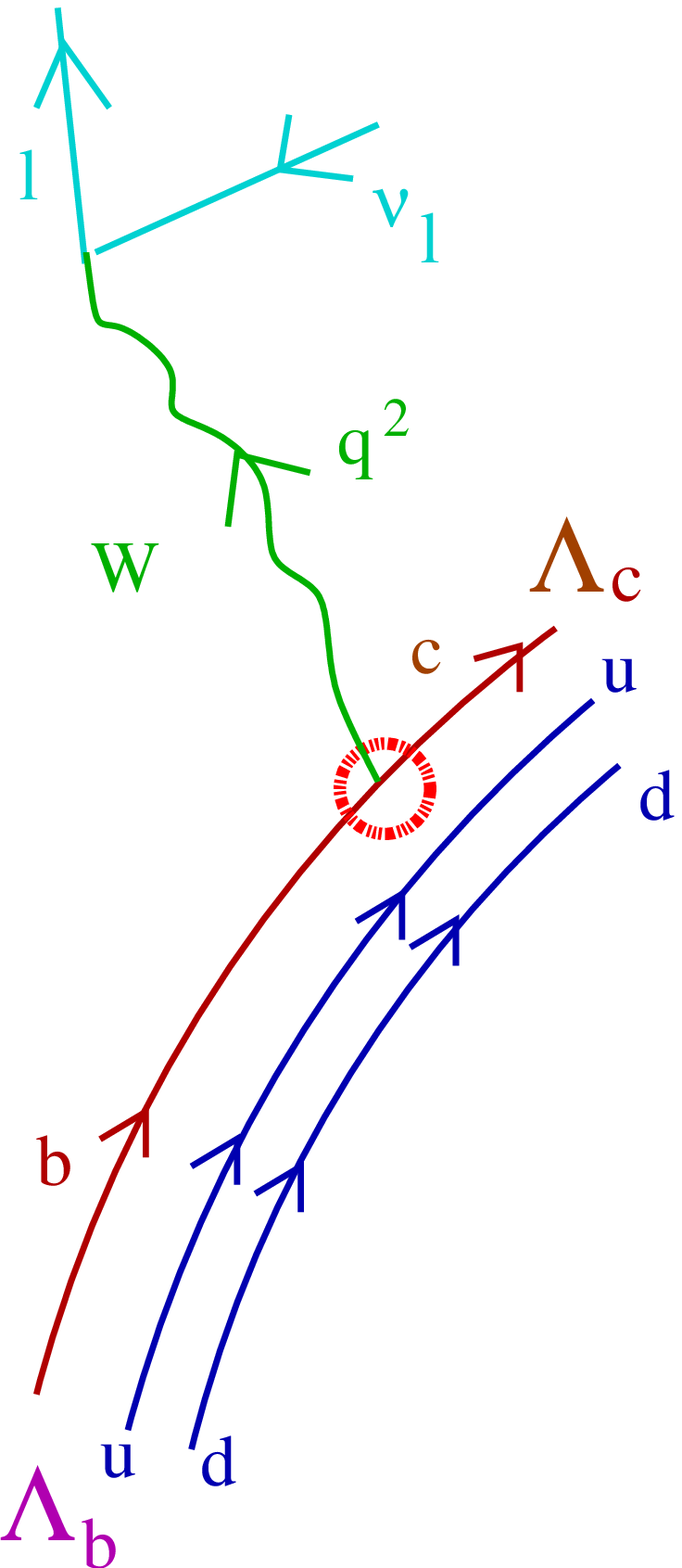}\hspace{1cm}\includegraphics[width=0.35\textwidth]{Xibccc_v2.eps}}
\end{center}
\caption{Definition of different coordinates (left) and sketch of the
  $\Lambda_b \to \Lambda_c$ decay (middle). Besides, in the right
  panel, taken from Ref.~\cite{Hernandez:2007qv}, we show the form
  factor combinations $(F_1 + F_2 +F_3)/\sqrt 2$ and $3G_1/\sqrt 8$ of
  the $\Xi_{bc} \to \Xi_{cc}$ transition (red), and $(F_1 +F_2
  +F_3)$ and $-\sqrt 3 G_1/\sqrt 2$ of the $\Xi^\prime_{bc} \to
  \Xi_{cc}$ transition (blue) evaluated using the AL1 interquark
  potential of Ref.~\cite{SilvestreBrac:1996bg}.}
\label{fig:coor}       
\end{figure}

In the Laboratory (LAB) frame (see left panel of Fig.~\ref{fig:coor}), the
Hamiltonian ($H$) of the three quark ($q,q^\prime, Q$, with $q,
q^\prime =l$ or $s$ and $Q=c$ or $b$) system reads:
\begin{eqnarray}
H&=&  \sum_{i=q,q^\prime,Q} \left (m_i
-\frac{\vec{\nabla}_{x_i}^2}{2m_i}\right ) +
V_{qq^\prime} + V_{Qq}+V_{Q q^\prime }
\end{eqnarray}
where $m_q,m_{q^\prime}$ and $m_Q$ are the quark masses, and the
quark-quark interaction terms, $V_{ij}$, depend on the quark
spin-flavor quantum numbers and the quark coordinates ($\vec{x}_1,
\vec{x}_2$ and $\vec{x}_h$ for the $q,q^\prime$ and $Q$ quarks
respectively). The nabla operators in the kinetic energy stand for
derivatives with respect to the spatial variables $\vec{x}_1, \vec{x}_2$
and $\vec{x}_h$. To separate the Center of Mass (CM) free motion,
we go to the heavy quark frame: $\vec{R},\vec{r}_1,\vec{r}_2$; 
 $\vec{R}$ and $\vec{r}_1$ ($\vec{r}_2$) are the CM position in
the LAB frame and the relative position of the quark $q$ ($q^\prime$)
with respect to the heavy quark $Q$.   The Hamiltonian now reads
\begin{eqnarray}
H&=&
-\frac{\vec\nabla_{\vec{R}}^2}{2 M} +
H^{\rm int} \\ H^{\rm
int}&=&-\sum_{i=1,2}\frac{\vec\nabla_i^2}{2\mu_i}-
\frac{\vec\nabla_1\cdot\vec\nabla_2}{m_Q}+
V_{qq^\prime}(\vec{r}_1-\vec{r}_2)+V_{Qq}(\vec
r_1)+V_{Qq^\prime}(\vec r_2)+ M
\end{eqnarray}
where $M = \left(m_q+m_{q^\prime}+m_Q\right)$, $\mu_{1,2} = \left (
1/m_{q,q^\prime} + 1/m_Q\right)^{-1}$ and $\vec\nabla_{1,2} =
\partial/\partial_{\vec{r}_1,\vec{r}_2}$. The intrinsic Hamiltonian
$H^{\rm int}$ describes the dynamics of the baryon, and it can be
rewritten as the sum of two single particle Hamiltonians ($h^{sp}_i$),
which describe the dynamics of the light quarks in the mean field
created by the heavy quark, plus the light--light interaction term,
which includes the Hughes-Eckart term ($\vec \nabla_1 \cdot
\vec\nabla_2 $).
\begin{eqnarray}
H^{\rm int} &=& \sum_{i=q,q^\prime} h^{sp}_i +
V_{qq^\prime}(\vec{r}_1-\vec{r}_2,spin) -
\frac{\vec\nabla_1\cdot\vec\nabla_2}{m_Q} + M  \\
&&h^{sp}_1 = -\frac{\vec\nabla_1^2}{2\mu_1} + V_{Qq}(\vec
r_1,spin),\qquad h^{sp}_2 = -\frac{\vec\nabla_2^2}{2\mu_2} + V_{Qq'}(\vec
r_2,spin)   \label{eq:defhsp}
\end{eqnarray}
In what respects to the quark--quark interactions, we use some
phenomenological ones~\cite{SilvestreBrac:1996bg} obtained from
quark--antiquark potentials fitted to a large sample of meson states
in every flavor sector\footnote{The usual $V_{ij}^{qq} = V_{ij}^{q
\bar q}/2$ prescription is assumed here.}. The
general structure is as follows ( $i,j=l,s,c,b$):
\begin{eqnarray}
V_{ij}^{q\bar q}(r) &=& -\frac{\kappa f_c(r)}{r}
+\lambda r^p - \Lambda + \Big\{\frac{a_0\kappa}{m_im_j}
\frac{e^{-r/r_0}}{rr_0^2}  +   \frac{2\pi\kappa^\prime f_c(r)}{3m_im_j}
 \frac{e^{-r^2/x_0^2}}{\pi^\frac32 x_0^3} \Big
\}\vec{\sigma}_i\vec{\sigma}_j,    \label{eq:phe}
\end{eqnarray}
with $\vec{\sigma}$ the spin Pauli matrices, $f_c(r) =  1 - e^{-r/r_c}$  and
$x_0(m_i,m_j)= A \left ( \frac{2m_im_j}{m_i+m_j} \right )^{-B}$.

In a baryon, the singlet color wave function is completely
anti-symmetric under the exchange of any of the three quarks. Within
the SU(3) quark model, we assume a complete symmetry of the wave
function under the exchange of the two light quarks ($u,d,s$) flavor,
spin and space degrees of freedom. On the other hand, for the
interactions described above, we have that both the total spin of the
baryon, ${\vec S}_{\rm B} = \left ( {\vec \sigma}_q + {\vec
\sigma}_{q^\prime} + {\vec \sigma}_Q \right )/2$, and the orbital
angular momentum of the light quarks with respect to $Q$, $\vec{L}~ [= 
{\vec l}_1 + {\vec l}_2$, with  ${\vec
l}_k=-{\rm i}~~{\vec r}_k \times {\vec \nabla}_k, \quad k=1,2$] 
commute with $H^{\rm int}$. We will assume that the ground
states of the baryons are in s--wave, $L=0$, which implies that the
spatial wave function can only depend on the relative distances $r_1$,
$r_2$ and $r_{12}=|\vec{r}_1-\vec{r}_2|$. Note that when the heavy
quark mass is infinity, the total spin of the light
degrees of freedom, $\vec{S}_{\rm light} = \left ( {\vec \sigma}_q +
{\vec \sigma}_{q^\prime} \right )/2$, commutes with $H^{\rm int}$, since the
$\vec{\sigma}_Q\cdot\vec{\sigma}_{q,q^\prime}/(m_Qm_{q,q^\prime})$
terms vanish in this limit. With all these ingredients, and taking
into account the quantum numbers of the light degrees of freedom for
each baryon, compiled in Table~\ref{tab:summ} and that in general are
always well defined in the static limit mentioned above, we have
constructed the wave functions in our variational approach. Thus, for
instance for $\Lambda-$type baryons ($\Lambda-$type baryons: $I=0,~S_{\rm light}=0$), we use\footnote{An
obvious notation has been used for the isospin--flavor ($|I,M_I\rangle
_I$, $|ls\rangle $ or $|sl\rangle $) and spin ($|S,M_S\rangle _{S_{\rm
light}}$) wave functions of the light degrees of freedom.}
\begin{eqnarray}
|\Lambda_Q; J=\frac12, M_J\left. \right \rangle  &=& \Big \{  |00 \rangle _I 
\otimes  |0 0 \rangle _{S_{\rm light}} \Big \} 
\Psi_{ll}^{\Lambda_Q} (r_1,r_2,r_{12}) \otimes  |Q; M_J\rangle  \label{eq:sim}
\end{eqnarray}
where $\Psi_{ll}^{\Lambda_Q} (r_1,r_2,r_{12}) = \Psi_{ll}^{\Lambda_Q}
(r_2,r_1,r_{12})$ to guaranty a complete symmetry of the wave function
under the exchange of the two light quarks ($u,d$) flavor, spin and
space degrees of freedom, and finally $M_J$ is the baryon total
angular momentum third component. Note, that SU(3) flavor symmetry (SU(2), in
the case of the $\Lambda_Q$ baryon) would also 
allow for a component in the wave function of the type
\begin{equation}
 \sum_{M_SM_Q} (\frac12 1 \frac12 | M_QM_SM_J)\Big \{  |00 \rangle _I 
\otimes  |1 M_S \rangle _{S_{\rm light}} \Big \} 
 \Theta_{ll}^{\Lambda_Q} (r_1,r_2,r_{12}) \otimes  |Q; M_Q\rangle  \label{eq:antis}
\end{equation}
with $\Theta_{ll}^{\Lambda_Q} (r_1,r_2,r_{12}) =
-\Theta_{ll}^{\Lambda_Q} (r_2,r_1,r_{12})$ (for instance terms of the
type $r_1-r_2$) and, the real numbers $(j_1j_2j|m_1m_2m)$ 
are Clebsh-Gordan coefficients. This
component is forbidden by HQSS in the limit $m_Q \to \infty$, where
$S_{\rm light}$ turns out to be well defined and set to zero for
$\Lambda_Q-$type baryons. The most general SU(2) $\Lambda_Q$ wave
function will involve a linear combination of the two components,
given in Eqs.~(\ref{eq:sim}) and (\ref{eq:antis}). Neglecting ${\cal
  O}(q/m_Q)$, HQSS imposes an additional constraint, which justifies
the use of a wave function of the type of that given in
Eq.~(\ref{eq:sim}) with the obvious simplification of the three body
problem. One can benefit~\cite{Albertus:2003sx} from similar
simplifications induced from HQSS for all the rest of baryons compiled
in Table~\ref{tab:summ}.
 
The spatial wave function, $\Psi_{qq^\prime}^{B_Q}$, is 
determined by the variational principle: 
\begin{equation}
\delta \langle B_Q | H^{\rm
int}| B_Q \rangle = 0
\end{equation}
For simplicity, we use  a Jastrow--type functional form
for the spatial wave function, as in the context of the similar
problem of  double $\Lambda$ hypernuclei~\cite{Al02}, 
\begin{eqnarray}
\Psi_{qq^\prime}^{B_Q} (r_1,r_2,r_{12}) &=& F^{B_Q}(r_{12})
\phi_q^Q(r_1)\phi_{q^\prime}^Q(r_2)
\end{eqnarray}
For simplicity, we do not entirely determine the
functions $\phi_q^Q$ and $\phi_{q^\prime}^Q$ from the variational
principle, but we rather fix the bulk of these functions to the
$s-$wave ground states ($\varphi_{i=q,q^\prime}^Q$) of the single
particle Hamiltonians, $h^{sp}_{i=q,q^\prime}$, defined in
Eq.~(\ref{eq:defhsp}), and modify their behavior at large
distances. Thus, we take
\begin{equation}
\phi_q^Q (r_1) = (1+\alpha_qr_1)\varphi_q^Q(r_1), \qquad
\phi_{q^\prime}^Q (r_2) = (1+\alpha_{q^\prime}r_2)\varphi_{q^\prime}^Q(r_2) 
\label{eq:onebody}
\end{equation}
with only one (two) free parameter for a $ll$ or $ss$ ($ls$) baryon
light quark content. Finally, we construct the light--light correlation
function, $F^{B_Q}$, from a linear combination of gaussians, with 
a total of eleven free parameters to be determined by the variational
principle. The mass of the baryon is just the expected value of the
intrinsic Hamiltonian.

Provided with this family of Jastrow type functions constrained by HQSS
and using several inter-quark interactions, we have calculated masses,
charge and mass radii of all bottom and charm baryons compiled in
Table~\ref{tab:summ} (see \cite{Albertus:2003sx}).  For the baryons
considered in \cite{SilvestreBrac:1996bg}, we agree remarkably well
with the results of this latter reference, obtained by solving
involved Faddeev equations, but thanks to HQSS, the baryon wave
functions are significantly simpler and more manageable than those
derived in \cite{SilvestreBrac:1996bg}.

Using the semi--analytical wave functions found here, we have also
studied the semileptonic decays~\cite{Albertus:2004wj} (see middle panel
of Fig.~\ref{fig:coor})
\begin{equation}
\Lambda_b^0 \to \Lambda_c^+ l^- {\bar \nu}_l, \qquad  \Xi_b^0 \to
\Xi_c^+ l^- {\bar \nu}_l 
\end{equation}
with $l=e,\mu$. We work on coordinate space, and develop a novel
expansion of the electroweak current operator, which supplemented with
heavy quark effective theory constraints, allows us to predict the
baryon form factors and the decay distributions for all $q^2$ values
accessible in the physical decays.  Our results for the partially
integrated longitudinal and transverse decay widths, in the vicinity
of the $q^2=(m_{\Lambda_b}-m_{\Lambda_c})^2$ point, are in excellent
agreement with lattice calculations~\cite{HB-Lattice}.  Comparison of
our integrated $\Lambda_b-$decay width to
experiment\cite{Amsler:2008zz,HB-exp} allows us to extract the
$V_{cb}$ Cabbibo-Kobayashi-Maskawa matrix element for which we
obtain~\cite{Albertus:2004wj} a value of
\begin{equation}
|V_{cb}| = 0.040\pm 0.005~({\rm
stat})~^{+0.001}_{-0.002}~({\rm theory})
\end{equation}
also in excellent agreement
with a recent determination, $|V_{cb}| = 0.0414 \pm 0.0012\,({\rm stat})\,\pm 0.0021\,({\rm syst})\,
 \pm 0.0018\,({\rm theory})$  from the
exclusive ${\bar {\rm B}^0_{\rm d}} \to {\rm D}^{*+}l^-{\bar \nu}_l$
decay~\cite{Exp3}.  Besides for the $\Lambda_b (\Xi_b)-$decay, the longitudinal and
transverse asymmetries, and the longitudinal to transverse decay ratio
are $\langle a_L \rangle=-0.954\pm 0.001~(-0.945\pm 0.002)$  , $\langle a_T
\rangle=-0.665\pm 0.002~(-0.628\pm 0.004)$ and $R_{L/T}=1.63\pm 0.02
~(1.53\pm 0.04)$, respectively. 

\section{Doubly Heavy Baryons}
We briefly review now static properties, semileptonic and
electromagnetic decays of doubly heavy baryons within the HQSS
constrained NRCQM scheme sketched in the previous section. We will
consider the ground states of baryons containing two charm, two bottom
or one charm and one bottom quarks (see Table~\ref{tab:sumhh}). In
these systems, HQSS amounts to the decoupling of the heavy quark spins
in the infinite heavy quark mass limit. In that limit one can consider
the total spin of the two heavy quark subsystem ($S_h$) to be well
defined~\cite{jenkins93}. To solve the baryon three--quark problem, we
will again use  a variational approach that leads to simple and
manageable wave functions thanks to the simplifications introduced in
the problem by considering $S_h$ to be well defined.
Details of this approach run in parallel to those described in the
previous section for single heavy baryons. Results for masses and
semileptonic $b\to c$ decay widths of these baryons, calculated with
different quark-quark interactions, can be found in
\cite{Albertus:2006ya}. We will focus here first, on the constraints
that HQSS imposes to the weak matrix elements that described these
semileptonic decays, and  second on the hyperfine mixing in the physical
$bc-$baryons, and the consequences of this mixing, in conjunction with
HQSS, for semileptonic and electromagnetic decays of the actual
$bc-$baryon states.

\begin{table}
\caption{Summary of the quantum numbers of the baryons containing two
  heavy quarks. $S_{hh'}^\pi$ stands for the spin parity of the heavy
  subsystem.}
\label{tab:sumhh}
\[ 
\begin{array}[t]{cllllc}
{\rm Baryon} & S & J^P & I & S_{hh}^\pi& {\rm Quark~ Content} \\[0.5ex]
\hline \\
 \Xi_{cc}      &  0 &\frac12^+ & \frac12 &  1^+& ccl\\
 \Xi^*_{cc}    &  0 &\frac32^+ & \frac12 &  1^+& ccl\\
\Omega_{cc}    & -1 &\frac12^+ & 0 & 1^+ & ccs\\
\Omega^*_{cc}  & -1 &\frac32^+ & 0 & 1^+ & ccs\\[1ex]
\Xi_{bb}       & 0  &\frac12^+ & \frac12 & 1^+ & bbl\\
\Xi^*_{bb}     &  0 &\frac32^+ & \frac12 & 1^+ & bbl\\
\Omega_{bb}    & -1 &\frac12^+ & 0 &  1^+ & bbs\\
\Omega^*_{bb}  & -1 &\frac32^+ & 0 &  1^+ & bbs
\end{array}
\qquad
\begin{array}[t]{cccccc}
{\rm Baryon} & S & J^P & I & S_{hh'}^\pi & {\rm Quark~ Content} \\[0.5ex]
\hline \\
\Xi'_{bc}      &  0 &  \frac12^+ & \frac12 & 0^+& bcl\\
\Xi_{bc}       &  0 &  \frac12^+ & \frac12 & 1^+& bcl\\
\Xi^*_{bc}     &  0 &  \frac32^+ & \frac12 & 1^+& bcl\\
\Omega'_{bc}   & -1 &  \frac12^+ & 0 & 0^+ & bcs\\
\Omega_{bc}    & -1 &  \frac12^+ & 0 & 1^+ & bcs\\
\Omega^*_{bc}  & -1 &  \frac32^+ & 0 & 1^+ & bcs
\end{array}
\]
\end{table}
Spin symmetry for both the $b$ and $c$ quarks enormously simplifies
the description of all semileptonic $\Xi_{bc}^{(\prime*)}\to
\Xi_{cc}^{(*)}\, l\, \bar \nu_l, \qquad \Omega_{bc}^{(\prime*)}\to
\Omega_{cc}^{(*)}\, l\, \bar \nu_l$ decays\footnote{Similar
conclusions hold for the $bb$ into $bc$ baryon decays.} in the limit
$m_{b,c} \gg \Lambda_\mathrm{QCD}$ and close to the zero recoil point
[$q^2=(m_{bc}-m_{cc})^2$]. All the weak transition matrix elements are
given in terms of a single universal
function~\cite{Flynn:2007qt}. Lorentz covariance alone allows a large
number of form factors (six to describe $\Xi_{bc}\to\Xi_{cc}$, another
six for $\Xi'_{bc}\to\Xi_{cc}$, eight each for
$\Xi_{bc}\to\Xi^*_{cc}$, $\Xi'_{bc}\to\Xi^*_{cc}$ and
$\Xi^*_{bc}\to\Xi_{cc}$, and even more for
$\Xi^*_{bc}\to\Xi^*_{cc}$). Let us consider, f.i., the
$\Xi_{bc}\to\Xi_{cc}$ and $\Xi'_{bc}\to\Xi_{cc}$ decays, in each case
the weak matrix element can be expressed in terms of three vector
($F's$) and three axial ($G's$) form factors ($r,r'$ are baryon
helicity indices, $u's$ are Dirac spinors, $v^\mu =
p^\mu/m_{\Xi_{bc}^{(\prime)}}$ and $v^{\prime\mu}
=p^{\prime\mu}/m_{\Xi_{cc}}$ are initial and final baryon velocities),
\begin{eqnarray}
\left\langle \Xi_{cc}, r'\ \vec{p}^{\,\prime}\left|\,
\overline c\gamma^\mu(1-\gamma_5)b(0)
 \right| \Xi_{bc}^{(\prime)}, r\ \vec{p}
\right\rangle&=& {\bar u}^{\Xi_{cc}}_{r'}(\vec{p}^{\,\prime})\Big\{ 
\gamma^\mu\left(F_1(w)-\gamma_5 G_1(w)\right)\nonumber\\
 && \hspace{-6cm} +v^\mu\left(F_2(w)-\gamma_5  G_2(w)\right) +v^{\prime\,\mu}\left(F_3(w)-\gamma_5  G_3(w)
\right)\Big\}u^{\Xi_{bc}^{(\prime)}}_r(\vec{p}\,), \quad w=v\cdot v'
\end{eqnarray}
In the $m_Q\to \infty$ limit, the above 12 form factors are not
independent and are all related to
a  single universal function. Finite $c$ and $b$ quark mass corrections turn out
to be small, as it can be appreciated in the right panel of
Fig.~\ref{fig:coor} (note that there, some combinations of form
factors that, in the $m_Q\to \infty$ limit,
should be equal or vanish~\cite{Flynn:2007qt} are plotted). An
extensive analysis of HQSS constraints on NRCQM semileptonic form
factors and decay widths of doubly heavy baryons can be found in
\cite{Hernandez:2007qv}.

To end this section, we will discuss the hyperfine mixing for
$bc-$baryons. In Table~\ref{tab:sumhh} we showed the $J^\pi=\frac12^+$
ground states of these baryons classified so that $S_h$ is well
defined ($S_h$-basis). Due to the finite value of the heavy quark
masses, the hyperfine spin interaction ($\vec\sigma\cdot\vec\sigma$
term in Eq.~(\ref{eq:phe})) between the light quark and any of the
heavy quarks can admix both $S_h=0$ and $S_h=1$ spin components into
the wave function~\cite{Roberts:2007ni}. This mixing should be
negligible for $bb$ and $cc$ doubly heavy baryons as the antisymmetry
of the wave function would require radial excitations and/or higher
orbital angular momentum in the $S_h=0$ component. However, in the
$bc$ sector one expects the actual physical $\Xi$ $(\Omega)$ particles
to be admixtures of the $\Xi_{bc},\,\Xi'_{bc}$
($\Omega_{bc},\,\Omega'_{bc}$) states listed in
Table~\ref{tab:sumhh}. This requires to diagonalize the Hamiltonian,
calculated for instance in the $S_h-$basis. This is easily done and
details can be found in
\cite{Albertus:2009ww}. Qualitatively\footnote{We will focus only on
  the $\Xi-$type baryons. The discussion is similar for the
  $\Omega-$type states.}, the physical eigenstates
($\Xi_{bc}^{(1,2)}$) turn out be quite close to those ($\hat\Xi_{bc},
\hat\Xi_{bc}^\prime$) in which the light quark $q$ and the $c$ quark
are coupled to well defined total spin $S_{qc}=0,1$
($S_{qc}-$basis). Note that, the $S_h$ and $S_{qc}$ basis are related
by a trivial rotation,
\begin{equation}
\hat\Xi_{bc}= \frac{\sqrt3}{2} \Xi'_{bc}+
   \frac{1}{2}\Xi_{bc}, \qquad \hat\Xi'_{bc}= -\frac{1}{2} \Xi'_{bc}+
   \frac{\sqrt3}{2}\Xi_{bc}
\end{equation}
In the $S_{qc}-$basis, hyperfine mixing is always inversely
proportional to the $b$ quark mass, and it is thus much smaller than
for the $S_h-$basis case. Indeed, NRCQM's calculations show that
physical and $S_{qc}-$basis states
differ in just a rotation of around $4^{\rm
  o}$~\cite{Albertus:2009ww}. 

Masses (eigenvalues) are very insensitive to hyperfine mixing, since
the non-diagonal terms induced by the hyperfine spin interactions are
around one thousand times smaller than the diagonal ones. However, as
Roberts and Pervin~\cite{Roberts:2007ni} pointed out, this mixing
could greatly affect the decay widths of doubly heavy baryons. NRCQM
calculations confirmed~\cite{Albertus:2009ww,Roberts} this strong
dependence of the semileptonic $b\to c$ decay widths on the hyperfine
mixing. This is not surprising, and it can be easily understood since
the $b\to c$ semileptonic decay width for transitions involving the
$S_h$-basis $\Xi_{bc}$ state is very much different from the
corresponding one involving the $\Xi'_{bc}$ baryon. This is a
straightforward prediction of HQSS~\cite{Flynn:2007qt} and its
validity was corroborated in the context of NRCQM's in
\cite{Hernandez:2007qv}. Indeed, HQSS predictions might be also used
to experimentally obtain information on the mixing angle for $bc$
baryons in a model independent manner~\cite{Albertus:2009ww}. The idea
is to use HQSS to predict width ratios for physical states, for
instance,
\begin{equation}
R_1^{\rm phys.}=\frac{\Gamma( \Xi^{(2)}_{bc}\to \Xi^*_{cc})}{\Gamma(
    \Xi^{(1)}_{bc}\to \Xi^*_{cc})}\sim  \tan^2\theta + {\cal
  O}(\frac{m_q,\Lambda_{QCD}}{m_c}), 
\end{equation}
that will depend, in the $m_Q\to \infty$ limit, only on the small
rotation angle $\theta$ between the $S_{qc}-$ and the physical
states\footnote{Note that HQSS relates the weak transition matrix
  element of the processes in both numerator and denominator of the
  above ratio. Thus neglecting mass differences, the phase space
  integrals can be performed leading to a model independent prediction
  for the ratio of widths.}. Experimental data, when available, could
be used to extract information on the admixtures in the actual
physical states.

Flavor conserving one-photon transitions $\Xi^*_{bc}\to
\Xi^{(1)}_{bc}\gamma,\, \Xi^{(2)}_{bc}\gamma, \quad \Xi^{(1)}_{bc}\to
\Xi^{(2)}_{bc}\gamma $ depend also on the mixing
angle~\cite{Albertus:2010hi}. As in the case of $b\to c$ semileptonic
decays, there are large corrections to these electromagnetic decay
widths due to the hyperfine mixing. However, here next-to-leading
${\cal O} (1/m_Q)$ corrections turn out to be quite large [f.i., from
  phase space $\Gamma_{em}\propto (M_i-M_f)^3$, and because
  $M_i\approx M_f$, the widths are very sensitive to the actual baryon
  masses], and in this case, it will not be possible to determine,
relying only on HQSS, the actual hyperfine mixing matrix.

\section{Concluding Remarks}

We have discussed some constraints and the enormous simplifications
that HQSS imposes on single and doubly heavy baryons. Within a NRCQM
scheme, i) we have variationally computed baryon masses and wave
functions, and used these latter ones to calculate $b\to c$
semileptonic decays of these baryons, ii) we have studied the
hyperfine mixing for  $bc-$baryons and shown that it greatly
affects their electromagnetic and semileptonic decay widths, and iii)
we have discussed how such dependence, when compared to the HQSS
predictions, might be used  to experimentally extract
information on the admixtures in the actual physical $bc$ baryons, in
a model independent manner.

\begin{acknowledgments}
  This research was supported by DGI and FEDER funds, under contracts
  FIS2008-01143/FIS, FIS2006-03438, FPA2007-65748, and the Spanish
  Consolider-Ingenio 2010 Programme CPAN (CSD2007-00042), by Junta de
  Castilla y Le\'on under contracts SA016A07 and GR12, and by the EU
  HadronPhysics2 project, grant agreement n. 227431. 
\end{acknowledgments}

\end{document}